\documentclass[prc,amsmath,twocolumn,superscriptaddress]{revtex4} 

\usepackage{dcolumn}
\usepackage{graphicx}
\usepackage{color}

\def\beq{\begin{equation}}
\def\eeq{\end{equation}}
\def\bea{\begin{eqnarray}}
\def\eea{\end{eqnarray}}
\def\vd{\langle v_d q \rangle}

\renewcommand*{\eqref}[1]{Eq.~(\ref{eq:#1})}
\newcommand*{\eqlab}[1]{\label{eq:#1}}
\newcommand*{\figref}[1]{Fig.~\ref{fig:#1}}
\newcommand*{\figlab}[1]{\label{fig:#1}}

\newcommand*{\secref}[1]{Section~\ref{sec:#1}}
\newcommand*{\seclab}[1]{\label{sec:#1}}

\def\VYP#1#2#3{{\bf #1}, #3 (#2)}  

\def\PL#1#2#3{Phys.~Lett.~\VYP{#1}{#2}{#3}}

\newcommand{\etal}{\mbox{\textit et al.}}                       %

\newcommand{\Omit}[1]{}

\begin{document}

\title{A Macroscopic Description of Coherent Geo-Magnetic Radiation from Cosmic
 Ray Air Showers}

\author{O. Scholten}
\email{scholten@kvi.nl}
\affiliation{Kernfysisch Versneller Instituut, University of Groningen,
9747 AA, Groningen, The Netherlands}
\author{K. Werner}
\affiliation{SUBATECH, University of Nantes -- IN2P3/CNRS-- EMN,  Nantes, France}
\author{F. Rusydi}
\affiliation{Kernfysisch Versneller Instituut, University of Groningen,
9747 AA, Groningen, The Netherlands}

\begin{abstract}
We have developed a macroscopic description of coherent electro-magnetic
radiation from air showers initiated by ultra-high energy cosmic rays due to the
presence of the geo-magnetic field. This description offers a simple and direct
insight in the relation between the properties of the air shower and the
time-structure of the radio pulse.
\end{abstract}
\maketitle

\section{Introduction}

In recent years the interest in the use of radio detection for cosmic ray air
showers is increasing with the promising results obtained from recent
LOPES~\cite{Fal05,Ape06} and CODALEMA~\cite{Ard06} experiments. These experiments
have in turn triggered plans to install an extensive array of radio detectors at
the Pierre Auger Observatory~\cite{Ber07}. There is thus a growing interest the
link between the properties of the air shower and the time structure of the
emitted pulse. Already in the earliest works on radio emission from air
showers~\cite{Jel65,Por65,Kah66,All71}, the importance of coherent emission was
stressed. Two mechanisms, Cherenkov radiation and geo-magnetic radiation were
proposed as possibilities. In more recent work~\cite{Fal03,Sup03,Hue03}, the
picture of coherent synchrotron radiation from secondary shower electrons and
positrons gyrating in the Earth's magnetic field was proposed. Extensive results
on geo-synchrotron emission, based on realistic Monte-Carlo simulations of the
shower development, are given in~\cite{Hue05,Hue07}.

The primary motivation of this work is to improve on the understanding of the
relation between the measured pulse shape using radio receivers and the
properties of the air shower induced by a cosmic ray. Therefore, we performed
macroscopic calculations which allow, under simplifying conditions, to obtain a
simple analytic expression for the pulse shape. This analytic expression shows a
clear relation between the pulse shape and the shower profile.

The picture we use is very similar to that used in Ref.~\cite{Kah66} which we
refine by using a more realistic shower profile and where we calculate the
time-dependence of the pulse. The magnetic field of the Earth induces, by pulling
with the Lorentz force the electrons and positrons in opposite directions, a net
electric current in the electron-positron plasma. This plasma moves with almost
the velocity of light towards the Earth at the front end of the cosmic-ray air
shower. In our approach the collective aspect is emphasized by treating this
induced current as macroscopic. This differs from the approach of
Refs.~\cite{Hue05,Hue07} where the motion of individual particles is stressed
(microscopic approach). In both the macroscopic and the microscopic approach the
emission of the electromagnetic pulse is caused by moving charges in the Earth's
magnetic field. Therefore these two pictures should be regarded as presenting a
complementary view of the same physical phenomenon. There are however differences
in the predicted pulse shapes and we hope that by presenting this complementary
picture the understanding of radio emission from extensive air showers can be
improved.

In the introduction of \secref{model} the basic outline of our approach is
presented and the various aspects are detailed in the different subsections.
Starting from a very basic picture we present our results in \secref{results}.
Subsequently the effects on the pulse shape are investigated of finite lateral
extend, finite pancake thickness, and a realistic energy distribution of the
electrons and positrons in the air shower.

\section{The Formalism\seclab{model}}

When an UHE cosmic-ray particle enters the upper layers of the atmosphere, a
cascade of high-energy particles -- called a cosmic-ray air shower -- develops.
Due to the high velocities, most of the particles are concentrated in the
relatively thin shower front, which, for obvious reasons, is called the
'pancake'. The pancake, which for the present discussion is assumed to be charge
neutral, contains extremely large numbers of electrons and positrons. Near the
core of the shower this pancake has a typical thickness of a few meters and is
moving to the surface of the Earth with (almost) the velocity of light through
the magnetic field of the Earth. The Lorentz force on the charged particles
induces an acceleration of the particles in the $\hat{x}$ direction, which is
perpendicular to the magnetic field and the shower axis. However, due to the
frequent collision with the air-molecules, where the relatively small transverse
velocity is randomized, this acceleration, when averaged over all electrons,
rather translates into a drift velocity and thus an electric current in the
$\hat{x}$ direction. This picture is similar to what happens to electrons in a
copper wire. When a voltage is applied over the wire, the electrons undergo a
constant acceleration due to the electric force which is however compensated by
collisions with the copper atoms, resulting in a constant drift velocity and thus
a constant electric current. At the surface of the Earth, electromagnetic
radiation can be detected, which is due to this relatively constant electric
current moving with high velocity towards the Earth. The shape of the
electromagnetic pulse is principally determined by the (relatively slow)
variation in time of the magnitude of the current, combined with time retardation
effects.

To emphasize the basic principles, we confine ourselves to a rather simple
geometry where the cosmic shower moves straight towards the Earth's surface (the
$-\hat{z}$ direction, see \figref{geom}) with velocity $\vec{v}_s=-\beta_s c
\hat{z}$ where $\beta_s\approx 1$. The position of the shower front above the
Earth's surface is given by $z=-\beta_s c t$, where the front of the shower
reaches Earth at time $t=0$. The Earth's magnetic field (with magnitude $B_E$) is
parallel to the surface (in the $\hat{y}$ direction), $\vec{B}=B_E \hat{y}$. The
strength of the induced electric current depends on the distance $h$ from the
front of the shower and on the time $t$ in the shower development. The direction
of the current is in the $\hat{x}$ direction.  All quantities are measured in the
rest system of the observer who is at rest at the surface of the Earth.

\begin{figure}[h]
 \centerline{\includegraphics[height=5cm,bb=0 0 595 515,clip]{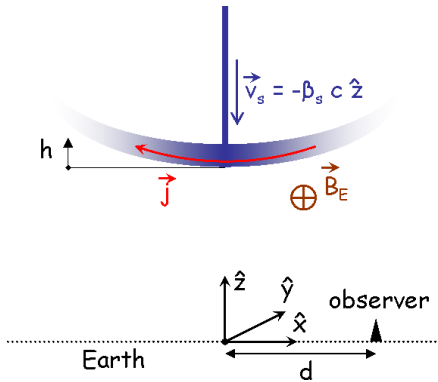}
 }
\caption[fig1-geff]{[color online]
\it The basic geometry as used in this paper. The pancake is shown as the shaded area,
moving with velocity $\vec{v}_s$. The observer is at a distance $d$ from the point of
impact. The curvature of the pancake has not been included in the present calculations.
}
  \figlab{geom}
\end{figure}

In the calculation of the current density we will initially assume a finite
extent in the horizontal directions (x and y). However, soon we will integrate
over these variables, knowing that the charged-particle density is strongly
peaked near the center of the shower. To emphasize the importance of the distance
$h$ behind the shower front, we will write the electron/positron density as
\beq
\rho_e(x,y,z,t)=\int  \tilde{\rho}_e(x,y,z,t,h) dh \;,
\eeq
where we assume a simple factorized form,
\bea
\tilde{\rho}_e(x,y,z,t,h) &=&N_e \delta(z+\beta_s ct+h) \nonumber\\
&\times& f_t(t) \rho_{\mbox{\tiny NKG}}(x,y) \rho_p(h) \;. \eqlab{rho-e}
\eea
The total number of charged particles at the time of maximum shower development
is denoted as $N_e$. The velocity of the shower front is given by
$\vec{v}_s=-\beta_s c \hat{z}$. The lateral distribution function is normalized
according to $\int \rho_{\mbox{\tiny NKG}}(x,y) \,dx\,dy=1$, the pancake
distribution obeys a similar normalization, $\int \rho_p(h) \, dh=1$, and the
maximum of the temporal (or longitudinal) distribution $f_t(t)$ is normalized to
unity. A detailed discussion of the parameterizations for these shower functions
is given in the appendix. In \secref{E-dis} also the effects of an energy spread
of the electrons and positrons are considered.

To emphasize the collective aspects of the model the calculation of the drift
velocity of the electrons and positrons is treated as a separate topic. The
magnitude of the induced current is calculated as the number of electrons (and
positrons) multiplied by an average drift velocity. In the following stage this
is combined with the shower profile to calculate the emitted electromagnetic
pulse.

\subsection{Magnitude of the Current\seclab{MagnCurr}}

For the present estimate it is assumed that there are equal numbers of positive
and negative charges moving towards the Earth with a large velocity. Due to the
Earth's magnetic field a net electrical current in the $\hat{x}$-direction is
induced with magnitude
\beq
 j(x,y,z,t)=\int  \vd \, e\, \tilde{\rho}_e(x,y,z,t,h) dh \;,
\eqlab{CurrDens}
\eeq
where $\tilde{\rho_e}$ is the density of electrons and positrons, \eqref{rho-e}.
To take into account that the electrons and positrons ($q=-1,+1$ respectively)
drift in opposite directions under the influence of the magnetic field, the
average sidewards drift velocity is weighted with the charge, denoted as $\vd$.

The radius of curvature of orbits of the electrons with an energy
$\epsilon_e=\gamma m c^2$ in the Earth's magnetic field is $R_B=\beta \gamma m
c/( e B_E )$. A realistic magnitude of the magnetic field
($B_E=0.3\times10^{-4}$~T) yields a curvature radius $R_B=\beta \gamma\times
50$~m. The angular deflection is thus $\theta=L/R_B$ where $L$ is the mean free
path, i.e.\ the length over which the electrons scatter over a large angle due to
multiple soft scattering or a hard scattering. The transverse component of the
velocity is $v_t=c \theta =c L e B_E/(\beta \gamma m c)$, assuming that
$\sin{\theta}\ll 1$ or a transverse velocity much smaller than the longitudinal
component. The  drift velocity, being the average over the complete trajectory,
is half this value~\cite{remark1},
\beq
v_d=c \theta/2 ={ c L e B_E\over 2 \beta \gamma m c}.
\eqlab{drift}
\eeq
The problem is thus now reduced to the calculation of the mean path length $L$.

At high energies, $\epsilon_e>10$~MeV, the electron cross section is dominated by
hard collisions and the mean path length is given by $L_R=X_0/\rho_{air}$, where
the electronic radiation length is $X_0=36.7$~g\,cm$^{-2}$ and the density of air
is $\rho_{air}=10^{-3}$~g\,cm$^{-3}$ at sea level. The density is of course lower
at higher altitudes.

\begin{figure}[h]
 \centerline{\includegraphics[height=6cm,bb=28 145 545 740,clip]{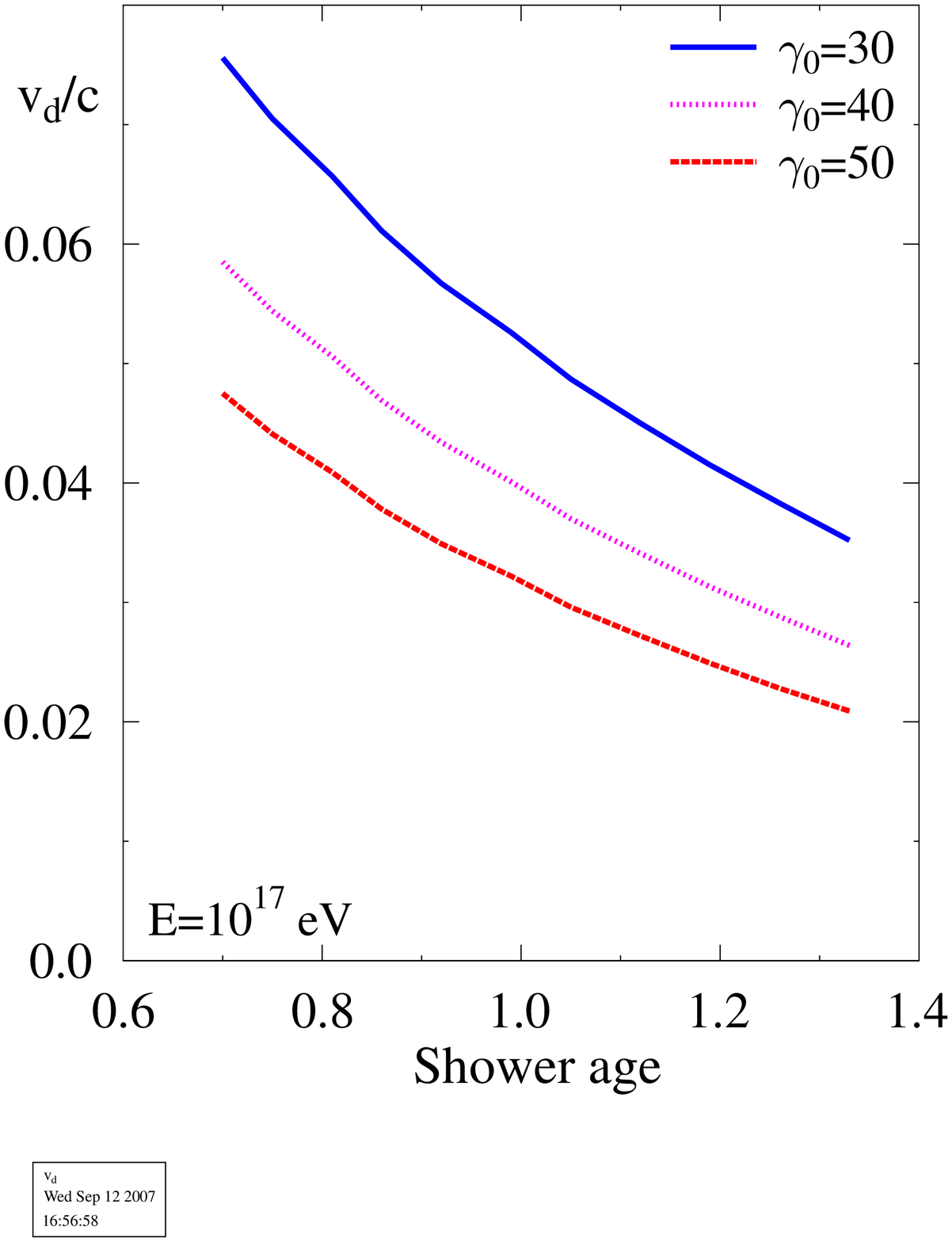}
}
\caption[fig17-eff]{[color online]
\it The drift velocity as function of shower age for different values of $\gamma_0$ in the
multiple Compton scattering contribution, \eqref{v-drift}. The range in shower age
corresponds to an elevation from 10~km up to sea level.}
  \figlab{v_d-dep}
\end{figure}

For smaller energies the above considerations do not apply since Coulomb
scattering becomes the dominant scattering mechanism with a mean free path
between successive collisions of 1~g\,cm$^{-2}$ (which can be determined directly
from the corresponding cross sections). Since Coulomb scattering is strongly
forward peaked, several collisions are necessary to randomize the momentum. As
argued in Ref.~\cite{Gai90} the scattering angle due to multiple Coulomb
scattering after traversing a distance $d$, is $\delta\langle\theta^2\rangle =d
(\gamma_0/\gamma)^2 \rho_{air}/X_0$ with $\gamma_0=40$. When
$\delta\langle\theta^2\rangle \approx 1$, the original direction is lost, giving
an effective path length $L_C=(\gamma/\gamma_0)^2 X_0/\rho_{air}$. Over the whole
energy range, the average path length $L$ may be written as $1/L=1/L_C + 1/L_R$
or
\beq
L={\gamma^2\over\gamma^2 + \gamma_0^2} {X_0 \over \rho_{air}} \;.
\eqlab{llldrift}
\eeq
The drift velocity, obtained from Eqs.(\ref{eq:drift},\ref{eq:llldrift}), now
reads
\beq
v_d= {c\over 2\,\beta } {\gamma\over\gamma^2 + \gamma_0^2} { e  B_E\over mc}
 {X_0 \over \rho_{air}}
 \; ,
 \eqlab{v-drift}
\eeq
keeping the energy dependence of the average path length. The average drift
velocity is finally obtained by averaging over the energy distribution of
electrons in a cosmic-ray air shower, using a parametrization given in
Ref.~\cite{Ner06} (see also \eqref{E-dis}).
Since this velocity is small, our assumption $\sin{\theta}\ll 1$ is indeed valid.

The drift velocity depends rather strongly on the assumptions made in the
estimate of multiple Coulomb scattering, as can be seen from \figref{v_d-dep} by
choosing different values for $\gamma_0$. The height dependence in $\vd$ is due
to the change in $\rho_{air}$ and due to the fact that the energy distribution of
particles in the shower pancake depends on shower age~\cite{Ner06}. In the
present calculations we used a constant drift velocity, $\vd=0.04$~c, equal to
the value at the shower maximum for $\gamma_0=40$.

In \secref{results} the results obtained using the average current density is
compared to the one obtained by explicitly integrating the electric field
generated by the particles of different energies. The difference between the two
appears to be mainly a normalization of the field strength, while the pulse shape
is hardly affected.

\subsection{The vector Potential}

Given a current density, $j^\mu$, the vector potential can be obtained using the
Li\'{e}nard-Wiechert fields,
\beq
A^\mu(x)= {1\over 4 \pi \varepsilon_0} \int
 \left. {j^\mu\over R(1-\vec{\beta}\cdot \hat{n})}\right|_{\mbox{ret}} \,dh\;,
 \eqlab{L-W}
\eeq
for a source with an infinitesimally small lateral extension. We use the common
notation where $\hat{n}$ is a unit vector pointing from the source to the
observer and $R$ is the distance, both evaluated at retarded time. Assuming that
all particles move with the velocity of the shower front, the denominator in
\eqref{L-W} can be rewritten to give
\bea
{\cal D} &=& R(1-\vec{\beta_s}\cdot \hat{n})|_{\mbox{ret}}
 = {c \over n}(t-t_r) - n\beta_s(h-c\beta_s t_r) \nonumber \\
 &=& \sqrt{(-c\beta_s t +h)^2 + (1-\beta_s^2 n^2)d^2}
 \eqlab{denom} \;,
 \eea
using \eqref{t_ret} for the retarded time. \eqref{denom} is written for a general
medium with an index of refraction $n$ however all our calculations are done in
the limit $n=1$. The distance between the observer and the point of impact of the
core of the air shower is denoted by $d$, see \figref{geom}.

Since the current density has only an $\hat{x}$-component, the vector potential
will share this property,
\beq
A^x(t,d)= J \int dh {\rho_p(h) f_t(t_r) \over {\cal D}} \;,\eqlab{Ax}
\eeq
where $\cal D$ is defined in \eqref{denom} and where the current density is
assumed to be parameterized according to \eqref{CurrDens} with $J= \vd N_e e/ 4
\pi \varepsilon_0 c$. We use SI units where  ${e\over 4\pi\varepsilon_0}=
1.44\times 10^{-9} $~[Vm] to get $E$ in [V/m]. \Omit{The expression for the
retarded time $t_r$ is given in \eqref{t_ret}.} The expression for the vector
potential shown in \eqref{Ax} is the central equation in our derivation.

\subsubsection{Charge Conservation, Static Dipole}

At the point above the Earth's surface where the shower front passes, the charges
are being pulled apart by the Lorentz force. The air shower can thus be regarded
as a `zipper', pulling apart positive and negative charges at the point where it
passes, leaving behind an electric dipole distributed along the path of the air
shower. Since we have argued that the electric current, which is associated with
the separating of the charges, is driving the electromagnetic pulse, we should
also investigate the effects of the created dipole. This dipole radiates because
it is not constant in time. To estimate its magnitude and the induced radiation
field, we will assume that the pancake thickness is infinitely small. Please note
that this dipole differs from the dipole mentioned in~\cite{Kah66} which is
co-moving with the air shower.

For definiteness, we temporarily assume that the charges are homogeneously
distributed over a distance $w$ in the $\hat{x}$ direction (this assumption will
be relaxed at the end). For a shower front at an height $z$ this corresponds to a
line-charge density $N_e f_t(-z/\beta_s c)/w$.  Since the charges move sideways
with a velocity $\vd$, a  charge $\Delta q=\pm\Delta t\,\vd\,N_e f_t(-z/\beta_s
c)/w$  accumulates at $x=\pm w/2$  after a time $\Delta t$. We will assume that
this charge is at rest in the Earth's reference system and remains fixed at all
later times while in reality it will slowly diffuse.  Since the shower front
progresses with a velocity $\beta_s c$, vertical line-charge densities $\pm
\rho_0(z)$ are created a distance $w$ apart with
\beq
\rho_0(z)={\vd\,N_e f_t(-z/\beta_s c) \over c\beta_s w} \;.
\eeq
These charge densities at height $z$ give a contribution to the zeroth component
of the vector potential of magnitude
\bea
 \delta A^0 &=& {1\over 4 \pi \varepsilon_0} \rho_0(z)
 \Big({1\over R^+} - {1\over R^-} \Big)\nonumber \\
&=& {1\over 4 \pi \varepsilon_0} {w\, \rho_0(z) \,x \over R^3}  =
 J {x \over c\beta_s R^3 } f_t({-z \over \beta_s c}) \;,
\eqlab{delA0}
\eea
where we have introduced the distance $R=\sqrt{z^2+d^2}$ with $d^2=x^2+y^2$ and
$R^\pm=\sqrt{z^2+(x\mp w/2)^2+y^2}$. From \eqref{delA0} it is clear that the
assumption of a homogeneous line-charge density can be relaxed at this point. The
scalar potential is obtained by integrating upward from the shower front over
$z$,
\bea
A^0(t,d)&=& J {x \over c\beta_s}
 \int_{z_0}^\infty dz {f_t(-z/\beta_s c) \over R^3} \;, \eqlab{A0}
\eea
where at time $t_r$ the shower front has reached a height of $z_0=-\beta_s c
t_r$. The charges are now taken into account for the full development of the
shower. \Omit{ Note that the denominator in \eqref{A0} differs from that in
\eqref{Ax} since the dipole moment is at rest in the system of the observer.}  It
should, however, be noted that gauge condition, $\partial_\mu A^\mu$, is not
fulfilled since in the present simple model we have assumed that the charges
forming the dipole moment are at rest in the Earth's system, while before they
were moving with a vertical velocity $\beta_s c$. This sudden acceleration
introduces an additional bremsstrahlung contribution which is beyond the scope of
the present work. \Omit{which is an artefact of the present model, will not occur
in an approach where the charge distributions are taken from a realistic
Monte-Carlo calculation.}

\subsubsection{Moving Dipole}

In the pancake, by virtue of the induced current, there will also be an induced
electric dipole moving towards the Earth with the shower velocity. We will argue
here that this dipole will not generate a contribution to the pulse in the limit
used in this paper, $n \beta_s=1$.

Due to the action of the Lorentz force the electrons and positrons will be
displaced an average distance $s$. The contribution to the vector potential can
now be written as
\bea
\Delta^m A^0(t,d)&\propto& \Big( {1\over{\cal D}^+ }- {1\over{\cal D}^- }\Big)
 = s {\partial \over \partial x} {1\over{\cal D} }
\nonumber \\
 &=& s {x (1-\beta_s^2 n^2) \over{\cal D}^3 }
 \eea
using the same notation as introduced in \eqref{delA0} and calculate the
derivative from \eqref{denom}. This contribution vanishes in the limit  $n
\beta_s=1$ and will be ignored in the following.

\subsection{The Electric Field
\seclab{nc}}

The electric and magnetic fields can be derived from the vector potential in the
usual way,
\beq
\vec{E}(t,d)=-\partial_0 \vec{A}(t,d) \;, \eqlab{E-field}
\eeq
where we have ignored the zeroth component of the vector potential (see the
discussion at the end of this Section).

Since the vector potential \eqref{Ax} has only a component in the $\hat{x}$
direction this will give rise to an electric field in the same direction. The
emitted radiation is thus linearly polarized in the $\hat{x}$ direction, i.e.\
perpendicular to the shower axis and the magnetic field.

The upper limit of the integral over $h$ in \eqref{E-field} extends up to
infinity\Omit{. However, using \eqref{Ax} for the case of interest where
$1=\beta_s^2 n^2$, the upper limit extends only up to $c \beta_s t$} and  we
obtain
\bea
&&E_x(t,d)
 = -J {d\over dt} \int_0^{\infty} dh {\rho_p(h) f_t(t_r) \over {\cal D}}
 \nonumber \\
 &=& -J\int_0^{\infty} \!\! dh \, \rho_p(h) {d\over dt}{f_t(t_r) \over {\cal D}}
 \nonumber \\
 &=& -J\int_0^{\infty} \!\! dh \, {f_t(t_r) \over {\cal D}} c\beta_s {d\,\rho_p(h)\over dh}
 \nonumber \\
 &-& J\int_0^{\infty} \!\! dh \, \rho_p(h)
  \Big({d\over dt}+c\beta_s{d\over dh} \Big) {f_t(t_r) \over {\cal D}}
 \nonumber \\
 &-& \left.c\beta_s \, \rho_p(h) {f_t(t_r) \over {\cal D}} \right|_{h=0}
 \; .
\eea
The second term can be rewritten as
\bea
&&\Big({d\over dt}+c\beta_s{d\over dh} \Big) {f_t(t_r) \over {\cal D}}
 \nonumber \\ &=&
  {1 \over {\cal D}} \Big({dt_r\over dt}+c\beta_s{dt_r\over dh} \Big)
 {df_t(t_r)\over dt_r}
 = {df_t(t_r)\over dt_r}{1 \over {\cal D}} \;.
\eea
Using $\rho_p(h)=0$ for $h=0$, the expression for the electric field simplifies
to
\bea
 E_x(t,d)
 &=& -J\int_0^{\infty} dh \, {f_t(t_r) \over {\cal D}} \beta {d\,\rho_p(h)\over dh}
 \nonumber \\
 &&- J \int_0^{\infty} dh \, {\rho_p(h) \over {\cal D}} {df_t(t_r)\over dt_r}
 \; , \eqlab{E_full}
\eea
where one should be careful in evaluating the integral because of the
$1/\sqrt{c\beta_s t-h}$ pole in ${1 \over {\cal D}}$. To investigate the effect
of this pole, associated with Cherenkov emission, we have explicitly studied the
case $\beta_s n> 1$ for which the pole in ${1 \over {\cal D}}$ lies inside the
integration region. Only for unrealistically large values for the index of
refraction Cherenkov radiation is emitted by the electric current density. For
realistic values of $n$ this effect is too small to distinguish. Since we see
that our predicted pulse shapes for realistic values of $n$ and $n=1$ are
identical we have limited ourselves to the latter.

\subsubsection{Limiting Case}

To obtain a simple estimate for the emitted radiation one may take the limit
$\beta_s=1$ and $n=1$ and ignore the thickness of the pancake, giving
\beq
{\cal D}=c\beta_s t  + {\cal O}(1-\beta_s^2)
\approx ct
\eeq
and, for positive values of $t$,
\beq
ct_r={ct\over 1+\beta_s} - {d^2 \over 2 c\beta_s t}+ {\cal O}(1-\beta_s^2)
 \approx -{d^2 \over 2 c t}  \;,
 \eqlab{t-ret-app}
\eeq
which is large and negative since $d^2 \gg c^2t^2$. Interesting to note here is
that the earlier part of the signal ($t$ small and positive) contains the
information of the earlier parts (at higher altitude) of the air-shower
development ($t_r$ large and negative). The electric field can now be calculated,
using $\int dh \, \rho_p(h) h\approx 0$, $\int dh \, \rho_p(h)=1$
\bea
E_x(t,d)
 &=& -J{n^2 \sqrt{d^2+c^2\beta_s^2 t_r^2} \over c{\cal D}^2}  {df_t(t_r)\over dt_r}
 \nonumber \\
 &+& J f_t(t_r) {c\beta_s^2 t \over {\cal D}^3}
  \;.
 \eqlab{E-nc-app}
\eea
In the limit $\beta_s=1$ and $n=1$ \eqref{E-nc-app} can be simplified further to,
\beq E_x(t,d)
 \approx J {c^2 t_r^2 4 \over d^4} \left[ t_r {df_t(t_r)\over dt_r}
 + f_t(t_r) \right] \;.
\eqlab{E-appx}
\eeq
The limit $t\to 0$ should be taken with care since this limit corresponds to
large (and negative) retarded times, see \eqref{t-ret-app}, where the air shower
may not even have started. As a result \eqref{E-appx} produces a finite electric
field at all times. For small distances to the shower core, $d\approx ct$, the
approximations made in deriving the expression for the retarded time,
\eqref{t-ret-app}, are no longer valid and \eqref{E-appx} thus not applicable.

From \eqref{E-appx} it can be seen that the time structure of the pulse is,
independent of the distance, given by a rather simple function of the
longitudinal shower profile. If at distance $d$ the peak of the pulse occurs at
time $t_0$, at twice the distance, the signal peak occurs at a time $4\times t_0$
and the signal is four times as broad. From \eqref{E-appx} it can be seen that
the peak value of the field, occurring at the same retarded time, has decreased
by a factor $2^4$. It should be noted that the emitted radiation does not contain
a relativistic Lorentz $\gamma$ factor and therefore does not depend on the exact
velocity of the shower front, as long as it is close to $c$. The dependence on
the energy distribution of the particles in the shower pancake is only indirectly
through the dependence of the drift velocity.

\eqref{E-appx} shows that there is a direct relation between shower profile and
the pulse structure. The calculations presented in \secref{results} indicate that
for a realistic case this feature is smeared out due to the effects of the finite
extend of the pancake and the lateral distribution.

\subsubsection{Dipole Field}

To obtain an estimate of the effect of omitting the zeroth component of the
vector potential from our discussion the contribution due to the electric dipole,
\eqref{A0}, to the electric field is calculated as
\bea
 -{d A^0 \over dx}&=&\Delta E_x=
  J\left.{n f_t(-z/\beta_s c) x^2\over {\cal D} R^3}\right|_{z=z_0}
 \nonumber \\ &&
 + J\int_{z_0}^\infty dz \,{f_t(-z/\beta_s c)\over \beta_s} {z^2+y^2-2x^2\over R^5}
 \;, \eqlab{dEx}
\eea
using ${d\,z_0 \over dx}={n\beta_s x\over {\cal D}}$. Following a similar
approach we obtain for the other components
\bea
\Delta E_y &=&
  J\left.{n f_t(-z/\beta_s c) x\,y\over {\cal D} R^3}\right|_{z=z_0}
 \nonumber \\ &&
 + J\int_{z_0}^\infty \! dz \,{f_t(-z/\beta_s c)\over \beta_s} {-3x\,y\over R^5}
 \;,  \eqlab{dEy}\\
\Delta E_z &=&
  J\left.{n f_t(-z/\beta_s c) x\,z\over {\cal D} R^3}\right|_{z=z_0}
 \nonumber \\ &&
 - J\int_{z_0}^\infty \!dz \,{f_t(-z/\beta_s c)\over \beta_s} {3x\,z\over R^5}
 \; . \eqlab{dEz}
\eea

It is important to note that in Eqs.~(\ref{eq:dEx}-\ref{eq:dEz}) the distance $R$
appears in the denominator in stead of ${\cal D}$ as in \eqref{E_full}. The
reason for this is that the electric dipole is at rest in the frame of the
observer. Since $R \gg {\cal D}$ for the cases of practical interest, the
contribution of \eqref{dEx} is much smaller than that of \eqref{E_full} and thus
can safely be ignored.

\subsection{Azimuthal Distribution of Radiation Pattern}

As remarked before, the emitted electric field due to the induced electric
current, \eqref{E_full}, is linearly polarized in the $\hat{x}$-direction. Its
magnitude depends only on the distance to the shower core and, for a shower with
a cylindrical symmetry, has a perfect azimuthal symmetry around the point of
impact of the shower. This symmetry is broken by the fact that, due to the drift
velocity which is induced by the magnetic field, the distribution of charged
particles in the shower is somewhat more stretched in the $\hat{x}$-direction
than in the $\hat{y}$-direction. In addition, the field of the induced dipole,
\eqref{dEx} and \eqref{dEy}, does not have an azimuthal symmetry. The symmetry
breaking induced by these two effects is however small and will not be considered
further.

\section{Results\seclab{results}}

\begin{figure}[h]
 \centerline{\includegraphics[height=6cm,bb=20 135 480 800,clip]{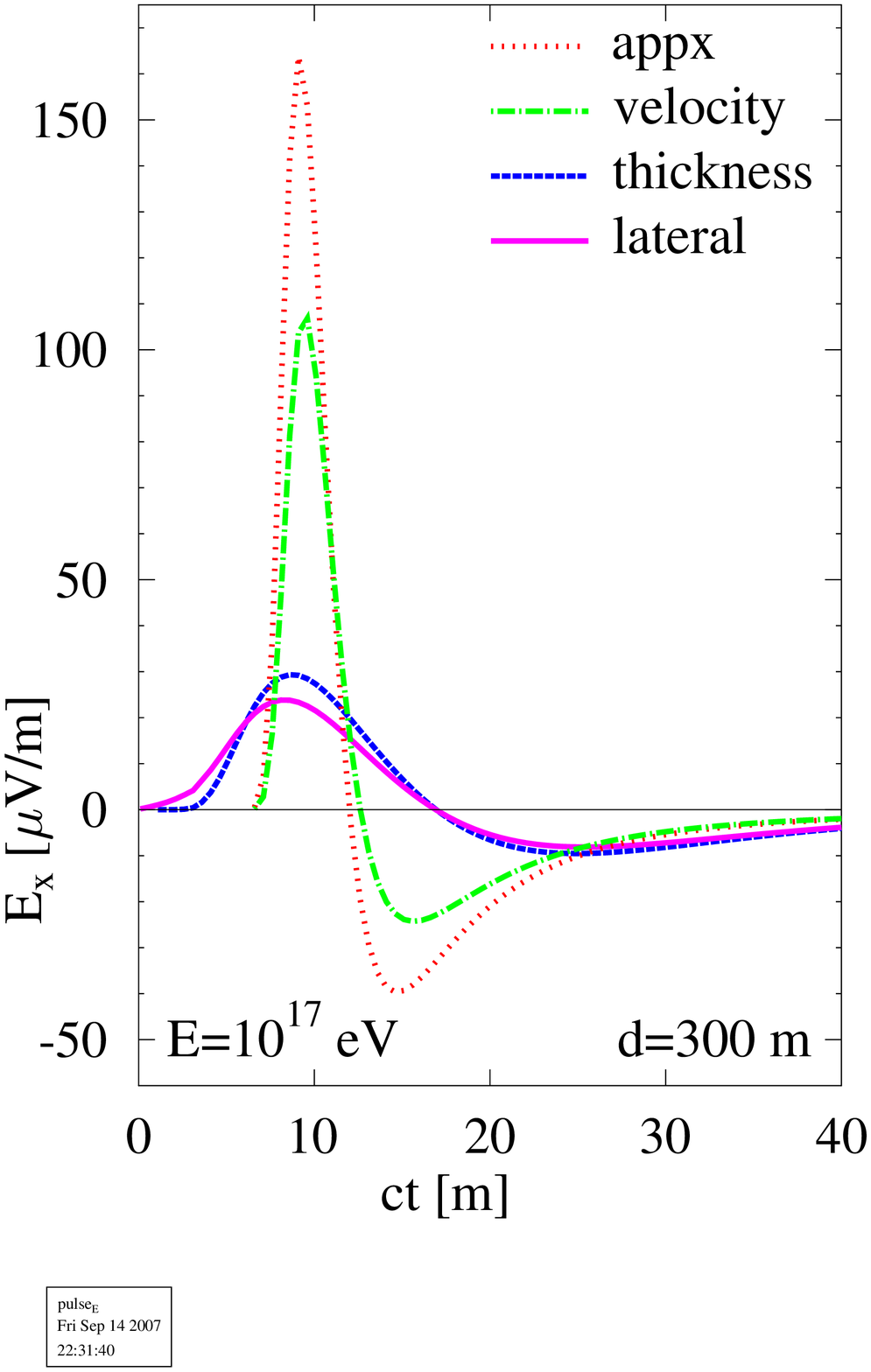}
 \
 \includegraphics[height=6cm,bb=20 135 480 800,clip]{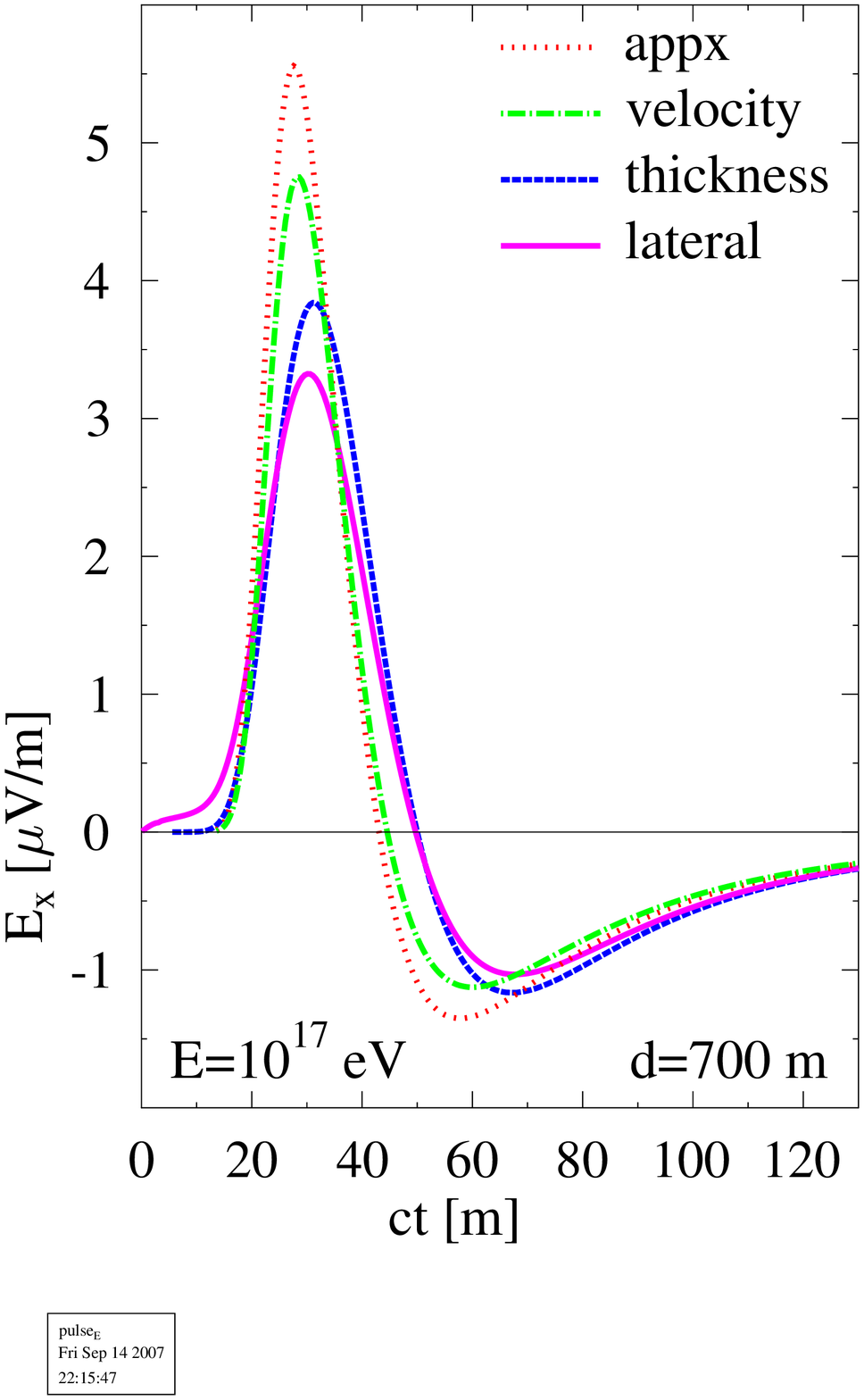}}
\caption[fig17-eff]{[color online]
\it Effects of refinements of the shower structure on pulse shape at $300$~m and
$700$~m from the shower core for a $10^{17}$~eV shower. The line labeled 'appx'
corresponds to the limiting case of \eqref{E-appx} while the other curves
include the effects of the finite pancake thickness, the lateral extent of the
shower and the energy distribution of electrons in the shower as discussed in the
text.}
  \figlab{Pulse-effects-t17}
\end{figure}

In \figref{Pulse-effects-t17} the calculated pulses as a function of time are
shown at distances of 300~m and 700~m from the shower core for different levels
of sophistication in the air-shower parametrization. The dotted curve, labeled
`appx', is the result of the most simple calculation using \eqref{E-appx} where
the longitudinal profile is given by \eqref{ft}. It has been verified that this
result is indistinguishable from that obtained using the full expression,
\eqref{E_full}, in the limit of vanishing pancake thickness. In order to
investigate the accurateness of this simple result as compared to a more
realistic calculation we relax some of the approximations to see their effects.

In arriving at \eqref{E_full}, the sideways drift velocity of the electrons and
positrons has been averaged over their energy distribution. In doing so, the
dependence of the denominator of \eqref{L-W} on electron energy has been ignored.
To test the effect of this approximation we have instead used the full
expression, see \eqref{A-E-dis}, resulting in the dash-dotted curves labeled
`velocity' in \figref{Pulse-effects-t17}, still in the limit of vanishing pancake
thickness. It shows that the effects of a spread in the energies of the electrons
and positrons in the pancake affects mainly the magnitude of the pulse and hardly
its time structure. At small distances the decrease of the peak height due to
this effect is stronger than at large distances.

Including a finite thickness of the pancake, using  \eqref{E_full} with
\eqref{rp}, results in the curves labeled `thickness' in
\figref{Pulse-effects-t17}. This has a very important effect on the pulse shape
at 300~m, which can easily be understood since a finie pancake thickness,
$L=c\Delta t$, introduces a `smearing' effect for the pulse over a time $\Delta
t$. Since for larger distances the pulse width is already sizable (it increases
roughly with the second power of the distance as follows from
Eqs.~(\ref{eq:t-ret-app}) and (\ref{eq:E-appx})) the smearing has only a minor
effect at 700~m.

Taking in to account the effects of the lateral spread of the particles in the
shower, \eqref{rNKG}, in addition to the pancake thickness, results in drawn
curve labeled `lateral' in \figref{Pulse-effects-t17}. The width of the pulse
increases even further although the effect is relatively small.

The effect of the static dipole field is shown in \figref{Pulse-dip}, where it is
compared with the pulse including the effects of a finite pancake thickness, for
two distances from the core. It should be noted that the dipole pulse has been
multiplied by a factor $10^3$ in order to be able to show it on the same scale.
This clearly shows the assertion made earlier that the dipole response can be
safely ignored. Apart from the small magnitude, also the associated long wave
length will make it undetectable in realistic experiments.

\begin{figure}[h]
\centerline{\includegraphics[height=6.0cm,bb=25 150 530 720 ,clip]{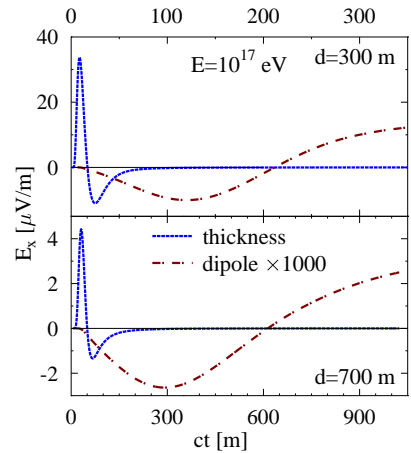}}
\caption[fig17-dip]{[color online]
\it Comparing the pulse due to the dipole field with that of the current at two distances
from the shower core.}
  \figlab{Pulse-dip}
\end{figure}

For a realistic shower one should expect a strong correlation between charged
particle velocity and the distance behind the shower front (slower particles
trailing further behind). For this reason we have chosen not to mix the effects
of the velocity distribution and finite pancake thickness in the present work
which is based on using simple parameterized showers. Results for a full
Monte-Carlo simulation will be presented in a future work, which will also take
into account the effects of an angular spread of the particles in the pancake.

\begin{figure}[h]
\centerline{\includegraphics[height=6.5cm,bb=20 135 533 805 ,clip]{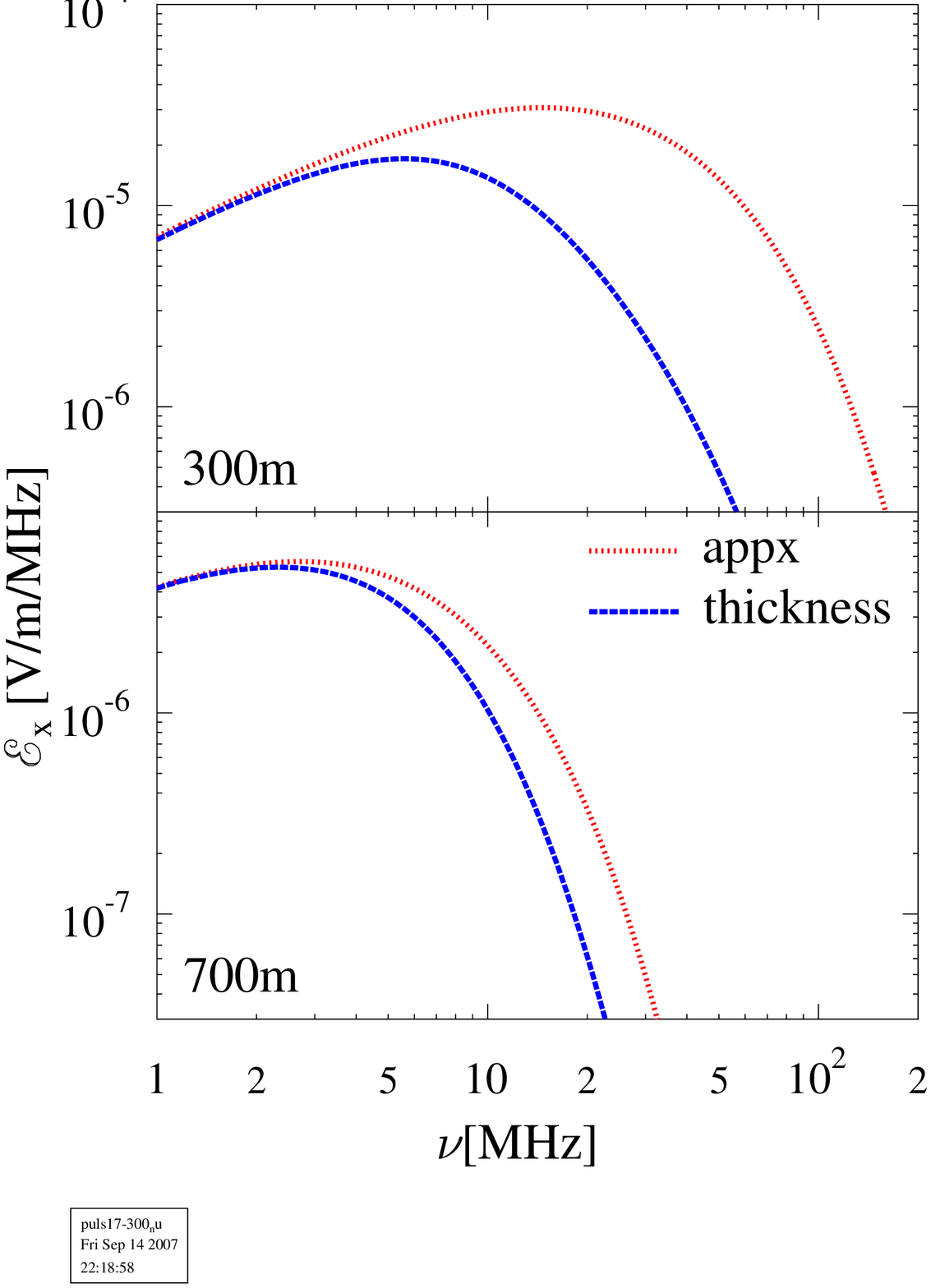}}
\caption[fig17-3n]{[color online]
\it Frequency response of the pulse at two distances from the shower core.
In the calculations labeled 'appx' and 'thickness' the thickness of the pancake
is zero (\eqref{E-appx}) and 10~m respectively.}
  \figlab{Pulse-n}
\end{figure}

The effect of a finite pancake thickness can also be seen in the frequency
response of the pulse as shown in \figref{Pulse-n}. The frequency response is
normalized such that $\int |E(t)|^2 dt=2\int |{\cal E}(\nu)|^2 d\nu$. At shorter
distances $d$, the effect of finite thickness is to suppress the higher frequency
components since the signal is only coherent for wave lengths larger than the
typical size of the emitting system. The signal, in the limit where the thickness
is ignored, does depend strongly on the distance from the shower core since the
projected longitudinal extend enters, which equals to zero when viewing the
shower head-on. Including a finite thickness reduces the dependence of the pulse
shape on the distance from the core.

\begin{figure}[h]
 \centerline{\includegraphics[height=7cm,bb=20 135 480 800,clip]{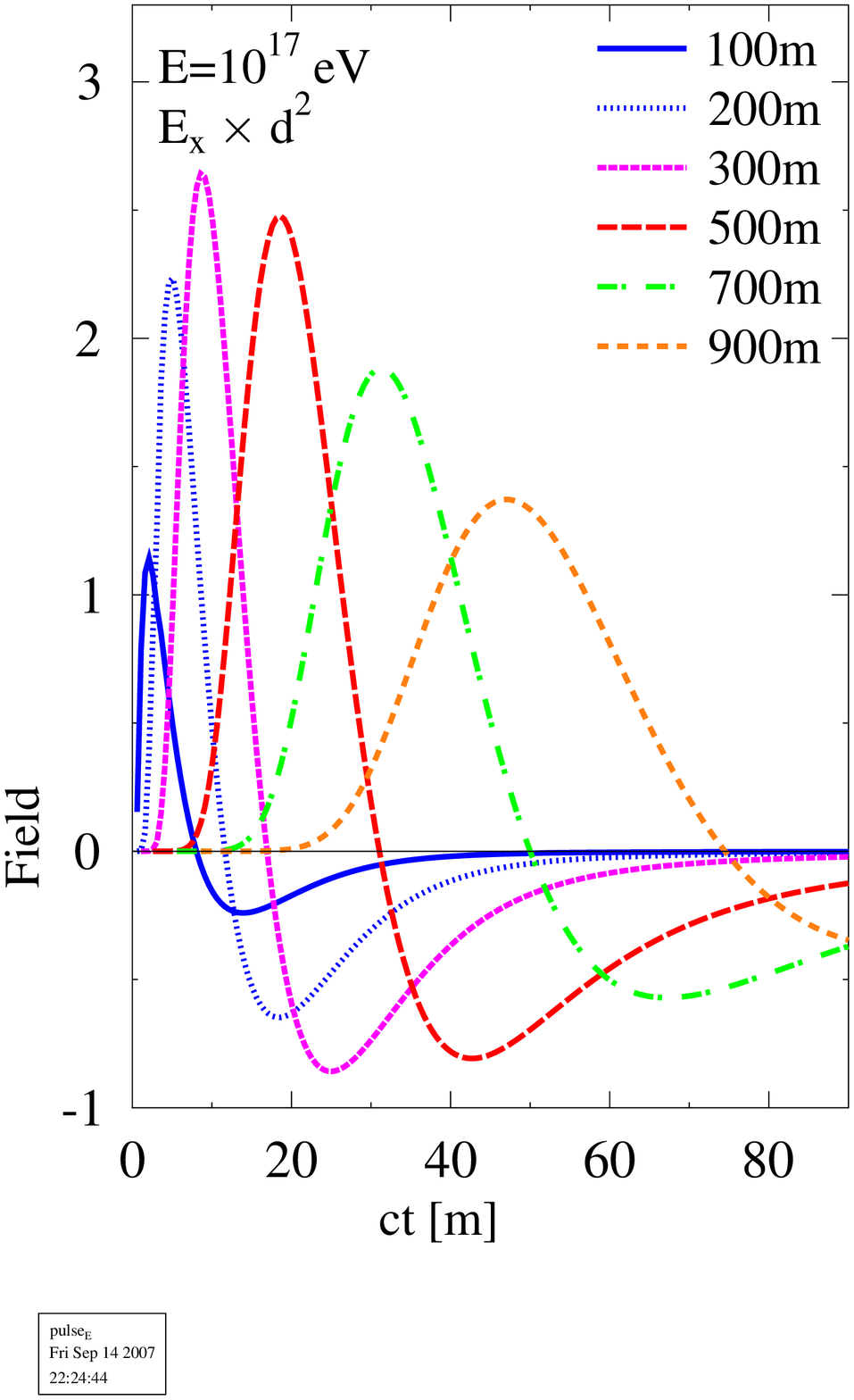}}
\caption[fig17-t]{[color online] \it Electric field strength (multiplied by $d^2$)
at different distances from the shower core as function of time.}
  \figlab{Pulse}
\end{figure}

In \figref{Pulse} the electric field is plotted as function of time for an
observer at various distances $d$ from the shower core. The primary energy is
$10^{17}$~eV and the calculation includes the effects of the pancake thickness
only. The shower core hits the Earth's surface at $t=0$. At large distances the
pulse decreases in magnitude even faster than $d^2$, as predicted by
\eqref{E-appx}. At small distances important deviations from the simple
parametrization are observed. This is to a minor extent due to the fact that the
approximations made to arrive at \eqref{E-appx} are no longer valid, and mostly
due to the effects of taking into account the finite thickness of the pancake
which strongly influences the pulse shape at distances $d<500$~m. Including
lateral extent of the shower will not greatly alter the picture.

\begin{figure}[h]
 \centerline{\includegraphics[height=6cm,bb=25 150 440 575,clip]{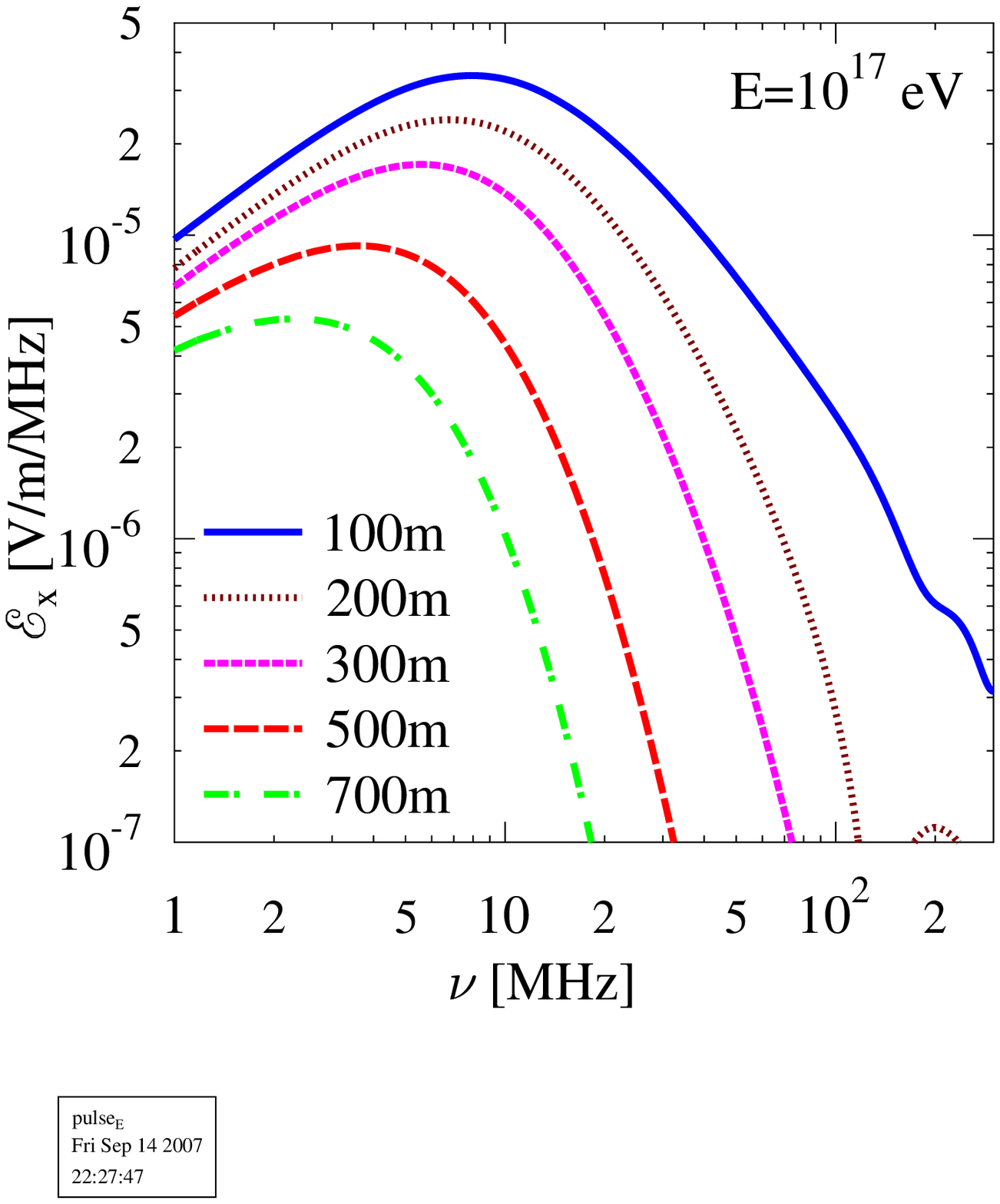}}
\caption[fig17-nu]{[color online] \it Fourier components of the electric field strength
at different distances from the shower core for a $10^{17}$~eV shower.}
  \figlab{Pulse-nu17}
\end{figure}

\begin{figure}[h]
 \centerline{\includegraphics[height=6cm,bb=25 150 440 575,clip]{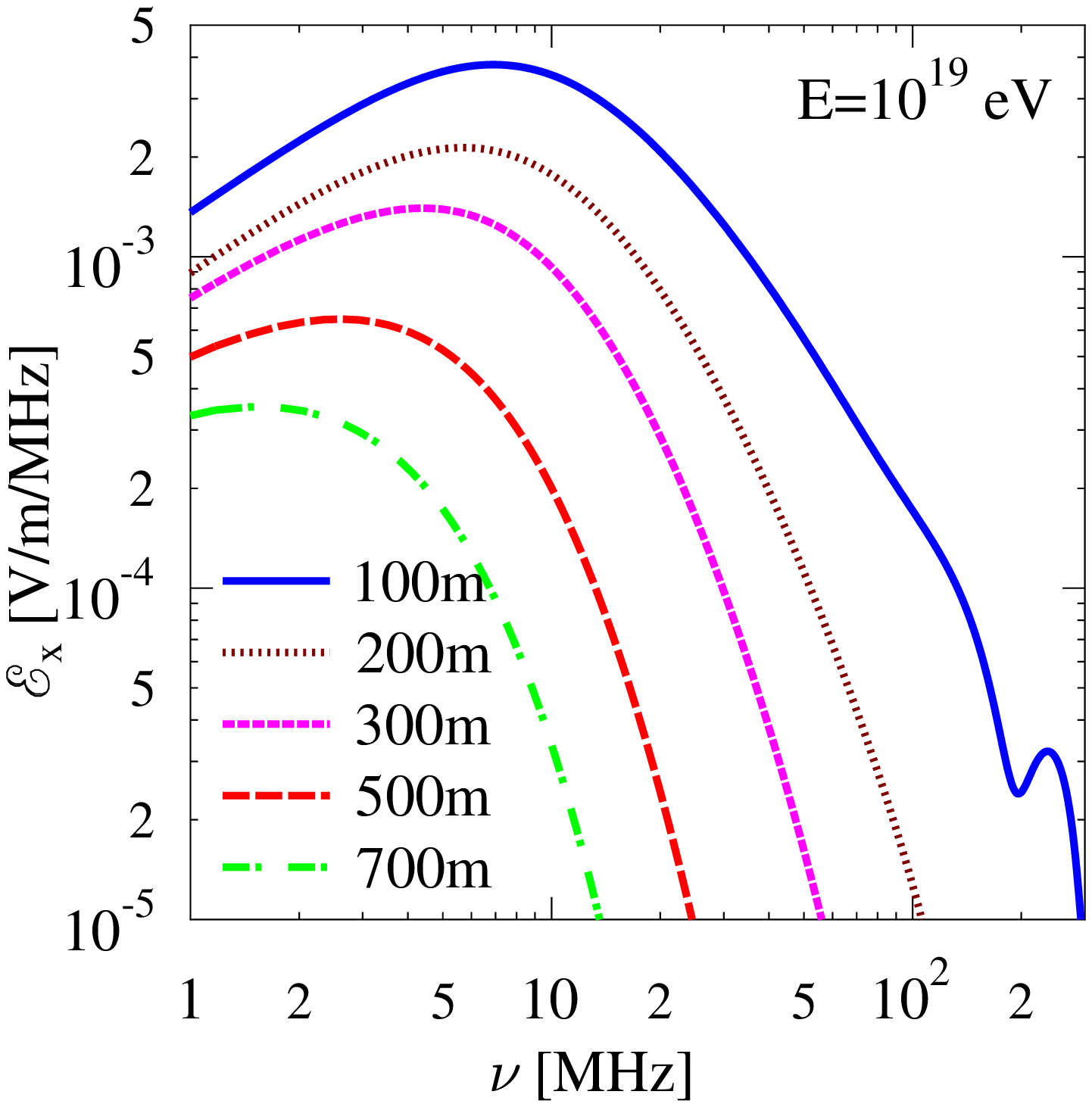}}
\caption[fig19-nu]{[color online] \it Same as \figref{Pulse-nu17} for a $10^{19}$~eV shower.}
  \figlab{Pulse-nu19}
\end{figure}

The frequency decomposition of the pulses  shown in \figref{Pulse} are shown in
\figref{Pulse-nu17} and those for an energy of $10^{19}$~eV in
\figref{Pulse-nu19}. At higher energies the shower maximum is closer to the
surface of the Earth making for a broader pulse. This is reflected in the
frequency spectrum by a peaking of the response at lower frequencies. At the same
time the number of charged particles in the shower is roughly proportional to the
energy of the primary particle. This in turn implies that the electric field is
two orders of magnitude larger for a $10^{19}$~eV induced shower than for
$10^{17}$~eV.

The present results can be compared with those given in Refs.~\cite{Hue05,Hue07}.
The basic features and magnitudes of the frequency responses are very similar. A
difference is seen in the time structure of the peak. The pulse form obtained in
this work has a distinct bi-polar structure, as has been discussed before, while
that of Refs.~\cite{Hue05,Hue07} has a simple unipolar structure. The bi-polar
structure of the pulse can also be understood from the fact that the vector
potential is positive definite and vanishes for both small and large times. The
electric field is the time derivative of this vector potential and crosses zero
at the time when the vector potential reaches a maximum. For the first part of
the pulse the two terms in \eqref{E-appx} add constructively resulting it the
large leading positive part. Since the vanishing vector potential at $t=0$ and
$t=\infty$ does not depend on details of the shower profile (radial distribution
or velocity distribution) the predicted bi-polar shape, with a vanishing
time-integral or zero-frequency component, can be regarded as a robust
prediction. This prediction also follows from the work of ref.~\cite{Kah66}. The
main difference between this work and ref.~\cite{Kah66} lies in the fact that we
have considered a more realistic shower profile and have presented a calculation
directly in the time domain. The latter allowed us to show explicitly the
relation between shower profile and pulse shape.

\section{Summary}

In this work a relatively simple macroscopic picture is presented for the
emission of coherent electromagnetic radiation from an extensive air shower
initiated by a high-energy cosmic ray. In this picture the radiation is emitted
through the electromagnetic current which is induced by the Earth's magnetic
field in the plasma at the front end of the shower. Under some simplifying
assumptions a simple algebraic equation can be derived, \eqref{E-appx}, which
clearly shows which are the important aspects of the air shower that determine
the electromagnetic pulse.

It is shown that the time structure of the pulse directly reflects the
longitudinal development of the number of electrons (and positrons) in the
shower. The radio pulse, therefore, gives very similar information on the air
shower as is obtained from air-fluorescence detection. Since the zero cross-over
point of the pulse is related to the maximum in the shower profile this implies
that the peak in the pulse is related to the shower development well before the
maximum.

In this first paper we have restricted ourselves to a very simple geometry. In a
future publication emission from showers for a more general geometry will be
investigated.

\begin{acknowledgments}
This work was performed as part of the research programs of the Stichting voor
Fundamenteel Onderzoek der Materie (FOM) with financial support from the
Nederlandse Organisatie voor Wetenschappelijk Onderzoek (NWO). We gratefully
acknowledge discussions with Stijn Buitink, Ralph Engel, Heino Falcke, Thierry
Gousset, Tim Huege, Andrey Konstantinov, and Sven Lav\`{e}bre on different
aspects of shower development and radio emission from air showers.
\end{acknowledgments}

\appendix

\section{Air-Shower Parametrization}

The front of the shower moves with a velocity $\beta_s c$ in the $-z$ direction,
towards the Earth were the point $z=0$ is taken at the Earth's surface. At time
$t=0$ the shower reaches the surface of the Earth. The shower thus exists at
negative times only.

The number of charged particles in the shower is parameterized as function of
time ($t$), distance from the shower front ($h$), and lateral distance
($r=\sqrt{x^2+y^2}$) where we consider here vertical showers only. For simplicity
we assume that the different dependencies simply factorize
\beq
\tilde{\rho}_e(x,y,z,t,h)=N_e f_t(t) \rho_{\mbox{\tiny NKG}}(x,y) \rho_p(h)
\delta(z+\beta_s ct+h) \;.\nonumber
\eeq
The distributions $\rho_p(h)$ and $\rho_{\mbox{\tiny NKG}}(r)$~\cite{NKG} are
normalized such that their integrals equal to unity. The total number of charged
particles at a specific time is thus given by $N_e f_t(t)$. This simple
parametrization of the shower is sufficient to gain insight into the basic
structure of the emitted electromagnetic pulse.

\subsection{Longitudinal Profile}

Following Ref.~\cite{Hue03}, the longitudinal shower development can be
parameterized using a shower age,
 \beq s(X)={3 X/X_0 \over X/X_0 + 2 X_{\mbox{\tiny mx}}/X_0} \;,
 \eeq
where $X_0$=36.7 g/cm$^2$ is the electronic radiation length in air. The primary
energy is denoted by $E_p$ and $X$ the penetration depth in units of
[g\,cm$^{-2}$]. The parameter $X_{\mbox{\tiny mx}}$ is chosen such as to
reproduce the positions of the shower maxima as have been determined from shower
simulations~\cite{Kna03},
 \beq
 X_{\mbox{\tiny mx}}=\Big( 840 + 70 \log_{10}(E_p/10^{20}\;eV) \Big)\;\mbox{g\,cm$^{-2}$} \;.
 \eeq
In the present calculations we model the atmosphere as
\beq
 X(z)=X(0) e^{-Cz} \eqlab{atmos}
 \eeq
with $X(0)=1000$~g\,cm$^{-2}$ and where $C$ 
is chosen such that $X(4\mbox{km})=630$~g\,cm$^{-2}$. Following Ref.~\cite{Hue03}
the time dependence of the number of charged particles is parameterized as
 \beq
N_e f_t(t)=N_e e^{(X-X_{\mbox{\tiny mx}} - 1.5 X\ln{s})/X_0} \;, \eqlab{ft}
 \eeq
using \eqref{atmos} with $z=-\beta_s t$. The maximum number of charged particles
is chosen as $N_e=6\times(E_p/10^{10}\;eV)$ to agree with the number given
in~\cite{Kna03} for a $10^{19}$~eV shower.

\subsection{Pancake Thickness}

In Ref.~\cite{Agn97} the measured arrival time distribution at a given radial
distance is fitted with a $\Gamma$-probability distribution function
($\Gamma$-pdf). Converted into a thickness of the shower front this can be
written as~\cite{Hue03}
\beq
 \rho_p(h)=h^\beta e^{-2h/L}\times (4/L^2)\;, \eqlab{rp}
\eeq
where the parameters $\beta$=1 and $L$ depend on shower age. At the shower
maximum a reasonable choice for the parameters is given by $\beta$=1 and
$L=10$~m. In the present calculation we keep these fixed for the full development
of the shower. The effects the curvature of the pancake has been ignored.

\subsection{Lateral Distribution}

The lateral particle density can be described with the NKG
(Nishimura-Kamata-Greisen)~\cite{NKG} parametrization, which at the shower
maximum ($s=1$) reads
\beq
\rho_{\mbox{\tiny NKG}}(x,y) = {1\over r^2_M} {2.5 \over 2 \pi}
\Big({ r \over r_M}\Big)^{-1} \Big({1 + {r\over r_M}}\Big)^{-3.5 } \;, \eqlab{rNKG}
\eeq
normalized such that $ 2\pi\,\int_0^\infty r\, dr\rho_{\mbox{\tiny NKG}}(x,y)=1$.
Here $r_M$ is the Moli\`{e}re radius at the atmospheric height of the maximum
derived from the atmospheric density as~\cite{Dov03} $ r_M(h) = {9.6 g\,
cm^{-2}\over\rho_{atm}(h) }$. The atmospheric density at a height of 4 km
corresponds to $\rho_{atm} = 0.82$~mg\,cm$^{-3}$, which in turn yields $r_M
\approx 117$~m.

\subsection{Energy Distribution\seclab{E-dis}}

In addition to the spatial distribution of the electrons in the shower, we have
also taken their spread in energy into account. All the time we assume that the
shower front is moving towards the Earth with velocity $\beta_s \approx 1$. The
distribution of electron energies ${\epsilon_e}$ in the shower is taken to be
$\rho_E(\epsilon_e)$, where the parametrization is taken from Ref.~\cite{Ner06},
\beq
{d \rho_E(\epsilon_e) \over d \log{\epsilon_e}}=
 {\cal N} {\epsilon_e \over (\epsilon_e+\alpha)(\epsilon_e+\beta)^s}\;.
 \eqlab{E-dis}
\eeq
This parametrization is based on a detailed comparison with the results of
Monte-Carlo calculations. The parameters are taken as~\cite{Ner06}
$\alpha=6.42522-1.53183 s$ and $\beta=168.168-42.1368 s$ in units of MeV where we
have implemented this parametrization for shower age $s=1$. The normalization
constant ${\cal N}$ is chosen to normalize the integral to unity.

The energy distribution of the electrons and positrons can be included in the
present calculation at two levels of sophistication. The simplest is to use it
only in the calculation of the average drift velocity as discussed in
\secref{MagnCurr}. In a somewhat more sophisticated approach the contribution to
the vector potential, \eqref{Ax}, is averaged resulting in
\beq
A^x=\int_0^{\beta_s} {\rho_E(\epsilon_e) v_d(\beta) \over R+\beta\beta_s t_rc}
N_e f_t(t_r) d\log{\epsilon_e} \;,
 \eqlab{A-E-dis}
\eeq
where $t_r$ is negative and the shower is at height $z_0=-\beta_s t_rc$.  The
denominator can be rewritten as $R+\beta\beta_s t_rc={\cal D} +z_0(\beta_s
-\beta)$. The shower front moves with velocity $\beta_s c$ while the charged
particles move with a smaller velocity and thus must be trailing behind the front
at a certain distance. The latter effect has not been included here. From the
vector potential the electric field can be calculated in the usual way.

\section{Retarded Time}

At the retarded time $t_r$ the front of the shower is at a height $z=-c\beta_s
t_r$. The travel time for a signal emitted at a distance $h$ behind the shower
front at $z$ to reach the observer (at Earth at distance $d$ from the point of
impact on Earth) is
\beq
t-t_r=\Delta t=n\sqrt{d^2 +(z+h)^2}/c \;,
\eeq
where $n$ is the index of refraction. The expression for the retarded time can
thus be written as
\beq
c\,t_r={c\,t-n^2 \beta_s h
 - n\sqrt{(-c\beta_s t +h)^2 + (1-\beta_s^2 n^2)d^2} \over 1-\beta_s^2 n^2 } \;
 \eqlab{t_ret}
\eeq
where in our convention the retarded time is negative for a shower above the
Earth and $t$ is assumed to be positive (the pulse, at a certain distance $d$,
arrives only after the shower has hit the ground).


\begin{thebibliography}{99}
\bibitem{remark1} In ref.~\cite{Kah66} the current is calculated using the
    transverse velocity at the end of the trajectory instead of the average which makes the
    current density a factor 2 larger in ref.~\cite{Kah66}.
\end{thebibliography}
\end{document}